\documentclass{iopart}

\usepackage{array}
\usepackage{graphicx}
\usepackage{bigstrut}
\usepackage{cite}
\usepackage{color}

\newcommand{\bk}{\mathbf{k}}
\newcommand{\bG}{\mathbf{G}}
\newcommand{\bbr}{\mathbf{r}}
\newcommand{\bR}{\mathbf{R}}
\newcommand{\bT}{\mathbf{T}}

\begin{document}

\title{Assessing the correlated electronic structure of lanthanum nickelates}

\author{Frank Lechermann}
\address{European XFEL, Holzkoppel 4, 22869 Schenefeld, Germany}
\address{Center for Computational Quantum Physics, 
The Flatiron Institute, 162 5th Avenue, New York, NY 10010, USA}

\begin{abstract}
The series of nickel-oxide compounds LaNiO$_3$ (formal Ni$(d^7)$), La$_2$NiO$_4$ (formal Ni$(d^8)$) 
and LaNiO$_2$ (formal Ni$(d^9)$) is investigated by first-principles many-body, using a combination
of density functional theory, self-interaction correction and dynamical mean-field theory. The
characteristics of these different nickelates, in good agreement with available
experimental data, is revealed by employing a compound-independent choice for the local Coulomb 
interactions. The dichotomy within the low-energy dominant Ni-$e_g$ sector of 
$\{d_{z^2},d_{x^2-y^2}\}$ kind is rising with growing Ni$(3d)$ filling across the series.
An intermediate-coupling scheme for spin-polarized calculations is introduced, which
leads to very weak Ni ordered moments for the infinite-layer compound LaNiO$_2$ in contrast
to the robust-moment system La$_2$NiO$_4$.

\end{abstract}

\maketitle

\section{Introduction}
Since the earliest days of studies on strongly correlated condensed matter~\cite{mot37},
research on nickel-oxide compounds has ever been an important aspect. But it just 
recently received a further boost when superconductivity was discovered 
in thin films of Sr-doped infinite-layer NdNiO$_2$~\cite{li19}. 
Nickelates are key examples of late transition-metal oxides where the $t_{2g}$ manifold
of the Ni$(3d)$ shell is in most cases filled, and the $e_g$ subshell plays the major
active role. While in neighboring (layered) cuprates that latter subshell physics 
reduces in the most interesting cases to an effective one-band/orbital picture~\cite{zha88}, 
the multiorbital character usually remains intact for nickelates. Rich phase diagrams with 
various competing orders are the result of this scenario (see e.g. Ref.~\cite{ima98} and 
references therein).

Concerning the research over the last 40 years, one may roughly group the respective 
dominant nickelate studies compound-wise into three areas. First, from traditional work on Mott 
insulators~\cite{mot37} and the apparent structural affinity to high-$T_c$ 
cuprates~\cite{bed86}, in the 1980s and early 1990s, the emphasis was on formal 
Ni$(3d^8)$ materials, e.g. NiO~\cite{saw84,huf84,laa86,elp92} and 
La$_2$NiO$_4$~\cite{goo82,say87,aep88,guo88,bat89,kui91}. Starting from the early 1990s
until later 2010s, main attention shifted towards formal Ni$(3d^7)$ compounds, 
mainly to the rare-earth (RE) perovskite(-like) RENiO$_3$ 
series~\cite{hor07}.
Though discussed early on~\cite{cre83}, work on formal Ni$(3d^9)$ systems heavily increased
after the finding of superconducting infinite-layer nickelates in the mid of 2019~\cite{li19}.

In this work we want to provide an theoretical overview over these different formal
Ni$(3d^{n=7,8,9})$ classes, by focussing on one representative material from each class and
describing them within an advanced first-principles many-body approach. Goal is to 
identify the key features of each class, to look for coherent theoretical settings
and similarities across the classes, and to weigh the quality of the description for
this wider nickelate family. To provide largest coherence from the ligand side, we
choose the following series of lanthanum nickelates for the investigation: 
LaNiO$_3$ (formal $d^7$), La$_2$NiO$_4$ (formal $d^8$) and LaNiO$_2$ (formal $d^9$).
While LaNiO$_3$ is well known as a paramagnetic metal down to lowest temperatures and
La$_2$NiO$_4$ as antiferromagnetic Mott/charge-transfer insulator over a wide temperature 
range, the definite electronic and magnetic characterization of LaNiO$_2$ is still a matter 
of intense debate. The role of an apparent self-doping band~\cite{nom19,wu19,lec20-1} 
in the latter compound complicates the correlation problem. 

It will be shown that for a rather coherent setting, our theoretical modelling delivers
good agreement with the key characterization and long-standing experimental results 
for LaNiO$_3$ and La$_2$NiO$_4$. For the same setting, we identify in the case of 
challenging LaNiO$_2$ an overall weak-metallic state (due to self doping) accompanied
by an orbital-selective Mott-insulating Ni-$e_g$ sector, very similar to recent findings 
for NdNiO$_2$.

\section{Methodology}

\subsection{General framework}
We utilize the DFT+sicDMFT scheme~\cite{lec19}, a charge self-consistent 
combination~\cite{sav01,pou07,gri12} of density functional theory (DFT), self-interaction 
correction (SIC) and dynamical mean-field theory (DMFT), to assess the
correlated electronic structure of LaNiO$_3$, La$_2$NiO$_4$ and LaNiO$_2$.
A mixed-basis pseudopotential (MBPP) code~\cite{els90,lec02,mbpp_code} takes care of the
DFT part in the local density approximation (LDA). To consider correlation
effects beyond LDA driven by the oxygen sites in the given late transition-metal oxides, 
Coulomb interactions on oxygen are described within SIC, and are incorporated in the  
O pseudopotential~\cite{vog96,fil03,kor10}. The SIC is applied to the O$(2s)$ 
and the O$(2p)$ orbitals via weight factors $w_p$ (see Ref.~\cite{kor10} 
for details). While the O$(2s)$ orbital is by default fully corrected with $w_p=1.0$, 
the adequate choice~\cite{kor10,lec19} $w_p=0.8$ is used for O$(2p)$ orbitals. 
The La$(4f)$ states are frozen in the pseudopotential core, as they supposed to be
empty and appear especially irrelevant for the key physics of the superconducting 
nickelates~\cite{zha20,osa21,zen21}. 
The full Ni$(3d)$ shell is used to construct the correlated subspace of the DMFT 
impurity problem via the projected-local orbitals formalism~\cite{ama08}. The 
projection is performed on the minimal set of Kohn-Sham (KS) states above the dominant 
O$(2s)$ bands, associated with O$(2p)$ and Ni$(3d)$. For LaNiO$_2$, the self-doping 
band is additionally included in the projection sphere~\cite{lec20-1,lec20-2,lec21}.
A five-orbital rotational-invariant Slater Hamiltonian, parametrized by a Hubbard $U$ and 
a Hund's exchange $J_{\rm H}$, is active in the correlated subspace. If not otherwise stated, 
we stick to $U=10$\,eV and $J_{\rm H}=1$\,eV as in previous nickelate 
studies~\cite{lec19,lec20-1,lec20-2,lec21}. One main goal of the present work is to check
wether such a compound-independent choice of local Coulomb parameters may satisfyingly
describe a wider range of different nickelates.

\subsection{Spin-polarized calculations}
The description of ordered magnetism in first-principles many-body calculations of DFT+DMFT
type leaves room for methodological interpretation~\cite{kes18}. As a hybrid scheme, the 
spin polarization can be handled in different ways. Allowing for local ordered spin moments 
in the DMFT part to investigate possible magnetically ordered phases is definitely 
adequate (e.g. Ref.~\cite{fle13}). There is 
common agreement that explicit correlation effects in partially-filled $d-$ or $f$-shells 
are a crucial ingredient of most spin-ordering phenomena in condensed matter. An additional
treatment also in the DFT part may have pros and cons. On the positive side, a strong
itinerant aspect of spin polarization, relevant for some weakly/moderately correlated 
compounds, might be better included. 
Also short-range 'ligand-coupling' effects could be more 
realistically covered by a DFT-based spin polarization of the charge density. 
On the negative side, the double-counting construction is not well suited for the explicit 
spin-polarization aspect. As a result, an often strong, static and temperature-independent 
exchange coupling within DFT dominates the physics, and subtle correlation effects in the 
magnetization are easily outshined. For instance, computation of Curie temperatures is more 
or less impossible with a full spin-polarized DFT+DMFT scheme. Since most DFT+DMFT
studies aim at correlations effect at strong coupling (i.e. close to a Mott-critical 
regime), therefore treating spin polarization only on the DMFT and not on the DFT level, 
has become the standard method for describing magnetically ordered phases.

In contrast to the both restrictive routes, i.e. DFT part spin-polarized or not, we here 
want to suggest an alternative way of handling spin polarization in the given context
of our nickelate study, which we term {\sl intermediate coupling (IC)} scheme. Key idea is to 
neglect only the local DFT part of spin 
polarization, since that one should for sure be comprehensively covered by the 
spin-polarized DMFT self-energies. The remaining strictly itinerant DFT charge-density
terms may still carry finite spin polarization.

In the mixed-basis pseudopotential approach~\cite{els90}, the KS wave function for Bloch 
vector $\bk$, band $\nu$ and spin $\sigma=\uparrow,\downarrow$ is expanded into plane waves 
(pw) and localized functions (lf), reading
\begin{equation}
\psi_{\bk\nu\sigma}(\bbr)=\frac{1}{\sqrt{\Omega_c}}\sum_{\bG}
\psi_{\bG}^{\bk\nu\sigma}\,{\rm e}^{i(\bk+\bG)\cdot\bbr}+
\sum_{\gamma lm}\beta_{\gamma lm}^{\bk\nu\sigma}\,\phi_{\gamma lm}^{\bk}(\bbr)\;,
\end{equation}
where $\Omega_c$ is the unit-cell volume, $\bG$ a reciprocal-lattice vector, 
$\gamma$ labels an atom in the unit cell and $lm$ are the usual angular-momentum 
quantum numbers. The localized functions are given by
\begin{eqnarray}
\phi_{\gamma lm}(\bbr)&=&i^l\,g_{\gamma l}(r)\,K_{lm}(\hat{\bbr})\;,\\
\phi_{\gamma lm}^{\bk}(\bbr)&=&
\sum_{\bT}{\rm e}^{i\bk\cdot(\bT+\bR_{\gamma})}
\phi_{\gamma lm}(\bbr-\bT-\bR_\gamma)\;,
\end{eqnarray}
whereby $g$ is a radial function, $K$ a cubic-harmonic function and $\bT$ a 
lattice vector.
Accordingly, the KS electronic charge density $\rho(\bbr)$ consists of three terms, i.e.
\begin{equation}
\rho_{\sigma}(\bbr)=\sum_{\bk\nu}f_{\bk\nu\sigma}|\psi_{\bk\nu\sigma}(\bbr)|^2=
\rho_{\sigma}^{\rm pw, pw}(\bbr)+\rho_{\sigma}^{\rm pw, lf}(\bbr)
+\rho_{\sigma}^{\rm lf, lf}(\bbr)\;.\label{ksterms}
\end{equation}
For instance, the purely-local third term $\rho^{\rm lf,lf}$ does not carry any
plane-wave part and reads
\begin{equation}
\rho^{\rm lf,lf}_{\sigma}(\bbr)=\sum_{\bk\nu\sigma}f_{\bk\nu\sigma}\left|
\sum_{\gamma lm}\beta_{\gamma lm}^{\bk\nu\sigma}\,\phi_{\gamma lm}^{\bk}(\bbr)\right|^2=
\sum_{\bT,\gamma lm}\rho^{\rm lf,lf}_{\gamma lm\sigma}(r')\,K_{lm}(\hat{\bbr}')\;,
\end{equation}
with $\bbr'=\bbr-\bT-\bR_\gamma$, and hence can be understood as an expansion into 
cubic harmonics. 

In the standard DFT+DMFT calculation for magnetically ordered phases of 
correlated materials, all terms in eq. (\ref{ksterms}) are spin-independent from 
a nonmagnetic KS treatment in each DFT step. Instead, we here suggest to perform
a spin-polarized KS calculation in each DFT step and the following representation of 
the associated charge density
\begin{equation}
\rho_{\sigma}^{\rm DFTpart}(\bbr)=
\rho_{\sigma}^{\rm pw, pw}(\bbr)+\rho_{\rm av}^{\rm pw, lf}(\bbr)
+\rho_{\rm av}^{\rm lf, lf}(\bbr)\;,\label{avterms}
\end{equation}
with
\begin{eqnarray}
\rho_{\rm av}^{\rm pw, lf}(\bbr)&=&
\frac{1}{2}\left(\rho_{\uparrow}^{\rm pw, lf}(\bbr)+
\rho_{\downarrow}^{\rm pw, lf}(\bbr)\right)\;,\\
\rho_{\rm av}^{\rm lf, lf}(\bbr)&=&
\frac{1}{2}\left((\rho_{\uparrow}^{\rm lf, lf}(\bbr)+
\rho_{\downarrow}^{\rm lf, lf}(\bbr)\right)\;.
\end{eqnarray}
Thus the fully local (lf,lf) term, as well as the semi-local (pw,lf) term are 
spin-averaged in the DFT part. Note that a coupling between local and plane-wave
part is already facilitated via the, then spin-dependent, projected-local orbitals 
$\sim \langle\phi_{\gamma lm}|\psi_{\bk\nu\sigma}\rangle$ in the construction of the 
correlated subspace~\cite{ama08}. It therefore proves indeed 
adequate to also average the (pw,lf) term in eq. (\ref{ksterms}), since the DMFT-relevant 
and therefore correlation-relevant intermixing between local and full Hilbert space
is already established at a different place in the complete DFT+DMFT 
framework~\cite{gri12}.

Using representation (\ref{avterms}) on the pure DFT level results of course in a
self-consistent vanishing of a possible ordered moment for correlated materials of 
interest. However in DFT+(sic)DMFT, the many-body part will reinject spin polarization
within the IC scheme, eventually with a more subtle and balanced linkage between the weakly 
and strongly correlated aspects of magnetic ordering.

\subsection{Crystal data and further computational settings}
All structural data for the La nickelates are taken from experiment. The LaNiO$_3$
compound crystallizes in a two-formula-unit cell with rhombohedral $R\bar{3}c$ space 
group, and we used the data by Zhang {\sl et al.}~\cite{zha17}. The structure is 
very similar to, but deviates slightly from a basic perovskite structure
(see top of Fig.~\ref{fig1}a). The Ni site is in an octahedral coordination and the
Ni-O bond length amounts to $d_{\rm Ni}^{\rm O}=1.929$\,\AA. The same coordination
holds also for Ni in La$_2$NiO$_4$, which is the single-layer $p=1$ compound of the
Ruddlesden-Popper series La$_{p+1}$Ni$_p$O$_{3p+1}$ and basically iso-structural to the 
famous high-$T_{\rm c}$ cuprate La$_2$CuO$_4$ as well as the famous low-$T$ superconducting 
ruthenate Sr$_2$RuO$_4$. At ambient temperature, La$_2$NiO$_4$ crystallizes in
the orthorhombic $Bmab$ structure (see mid of Fig.~\ref{fig1}a) with a
two-formula-unit cell. The in-plane Ni-O distance amounts to 
$d_{\rm Ni}^{\rm O,ip}=1.948$\,\AA, while the out-of-plane distance to the apical oxygen
reads $d_{\rm Ni}^{\rm O,oop}=2.267$\,\AA. Above $T\sim 770$\,K the compound 
transforms to the simpler tetragonal $I4/mmm$ structure. Below $T\sim 80$\,K the
system transforms to a new tetragonal phase with $P4_2/ncm$ symmetry, however that
phase is not considered in the present work. All lattice data on 
La$_2$NiO$_4$ is taken from the work by Rodriguez-Carvajal {\sl et al.}~\cite{rod91}.
Finally, the LaNiO$_2$ compound with $P4/mmm$ space group cannot be grown in 
single-crystal form, but exists as polycrystals~\cite{hay99} and in thin films on 
SrTiO$_3$~\cite{osa21,zen21}. This infinite-layer structure consists of NiO$_2$
layers with $d_{\rm Ni}^{\rm O}=1.980$\,\AA\, and missing apical oxygen 
(see bottom of Fig.~\ref{fig1}a). We use the lattice data for the single-phase 
model from Hayward {\sl et al.}~\cite{hay99}. The general increase of 
$d_{\rm Ni}^{\rm O}$ from LaNiO$_3$ to La$_2$NiO$_4$ and LaNiO$_2$ may be associated
with the expected lowering of the Ni oxidation state.

For the DFT+sicDMFT calculations we utilize a plane-wave cutoff $E_{\rm cut}=16$\,Ry
and a 13$\times$13$\times$13 $k$-point mesh for LaNiO$_2$. The respective 
$k$-point mesh for the other structure is accordingly tailored to yield the
identical $k$-point density. Localized functions in the MBPP formalism are introduced for 
La$(5d)$, Ni$(3d)$ as well as O$(2s,2p)$. Continuous-time quantum Monte Carlo in 
hybridzation expansion~\cite{wer06} as implemented in the TRIQS code~\cite{par15,set16} 
is utilized to solve the quantum-impurity problem. The fully-localized-limit 
double-counting scheme~\cite{ani93} is applied. Maximum-entropy~\cite{jar96} and 
Pad{\'e}~\cite{vid77} methods are employed for the analytical continuation from 
Matsubara space onto the real-frequency axis. Paramagnetic (PM) calculations are 
performed for LaNiO$_3$. In the case of La$_2$NiO$_4$ and LaNiO$_2$, both, paramagnetic and 
antiferromagnetic (AFM) phases are investigated, for details see the corresponding 
sections~\ref{113} and~\ref{214}.

\section{General electronic structure aspects}
\begin{figure}[t]
\begin{center}
\includegraphics*[width=12.75cm]{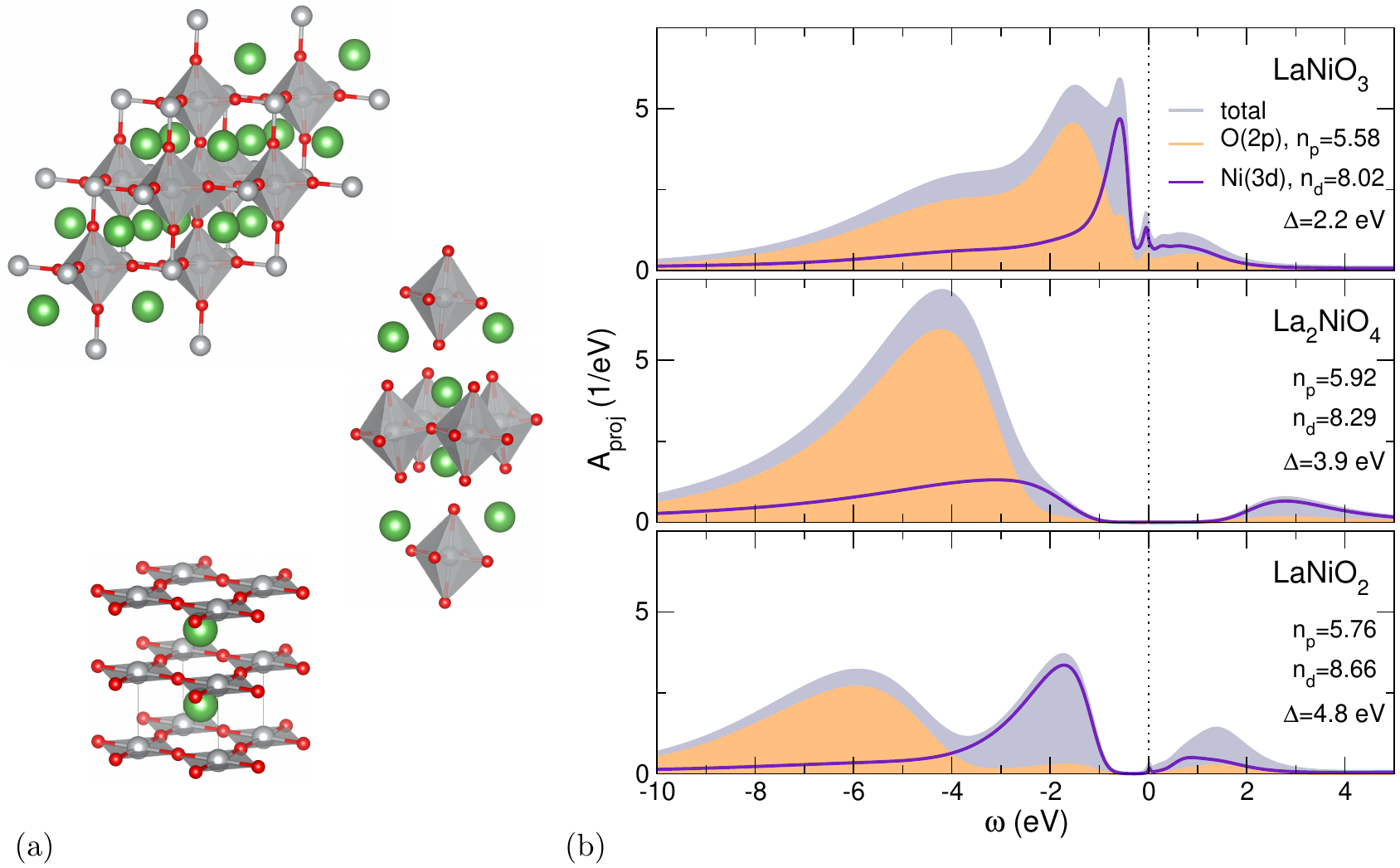}
\end{center}
\caption{Structural and spectral overview for the nickelate compounds  
LaNiO$_3$, La$_2$NiO$_4$ and LaNiO$_2$. 
(a) Top to bottom: ambient-temperature crystal structure of the compounds. 
LaNiO$_3$: rhombohedral $R\bar{3}c$; La$_2$NiO$_4$ orthorhombic $Bmab$; LaNiO$_2$:
tetragonal $P4/mmm$. La (green), Ni (grey) and oxygen (red).
(b) Top to bottom: DFT+sicDMFT total and projected spectral function for PM-LaNiO$_3$, 
AFM-La$_2$NiO$_4$ and PM-LaNiO$_2$ at $T=193$\,K, respectively. The values for
$n_p$ and $n_d$ refer to the O$(2p)$ and Ni$(3d)$ filling as obtained from integrating
$A_{\rm proj}(\omega)$. The respective charge-transfer energy $\Delta$ is computed
within DFT+sic (see text).\label{fig1}}
\end{figure}

Before discussing individual aspects of each of the studied nickelate compounds, let
us start with an overview of the correlated structure when lowering the
Ni 'oxidation potential' from LaNiO$_3$ via La$_2$NiO$_4$ to LaNiO$_2$. 
Figure~\ref{fig1}b displays from top to bottom the total and site-orbital-projected 
spectral function $A(\omega)$ for the three oxides. Note again that the local
Coulomb parameters $U=10$\,eV and $J_{\rm H}=1$\,eV are kept identical for each
compound.

The slightly-distorted perovskite LaNiO$_3$ is the metallic end member of the RENiO$_3$
series. Formally, nickel is in the $3+$ oxidation state, i.e. a local 
$3d^7(t_{2g}^6e_g^1)$ configuration would hold. Yet from the present calculations,
Ni is closer to a $3d^8(t_{2g}^6e_g^2)$ configuration with a ligand hole on oxygen, i.e.
a $3d^8\underline{L}$ state. 
In detail, according to the projected spectral function, the average O$(2p)$ filling 
per oxygen site $n_p=5.58$ translates into about 1.2 holes per unit cell. A significant
oxygen hole density can also easily be observed from substantial O$(2p)$ spectral
weight above the Fermi level $\varepsilon_{\rm F}^{\hfill}$. This finding is in line
with previous works~\cite{dem93,miz00,par12,lau13,joh14,sub15,bis16} that 
emphasize the preference for oxygen-hole formation in LaNiO$_3$ and presumably in further 
formal-$3d^7$ nickelates. Under these circumstances, the (effective) charge-transfer 
energy $\Delta_{\rm (eff)}$ should be small or even negative~\cite{ima98}. We here define 
$\Delta=\varepsilon_d-\varepsilon_p$ as obtained from DFT+sic, i.e. nonmagnetic
DFT calculations with SIC on oxygen, and $\varepsilon_{d,p}$ as the respective 
centers of the associated Ni$(3d)$ and O$(2p)$ projected-local-orbital bands. 
The resulting $\Delta_{\small\mbox{LaNiO$_3$}}=2.2\,$eV is indeed rather small.

The Ni$(3d)$ filling $n_d$ raises in La$_2$NiO$_4$ and the number of oxygen holes is
comparatively small. The Ni$^{2+}$ state with $3d^8$ configuration is indeed closest 
to reality. The system is an AFM insulator with a charge gap $\Delta_g$ of about 
$4$\,eV, in line with experiment~\cite{eis92}. However note that compared to prototypical 
NiO~\cite{lec19}, the valence-band-maximum states are of stronger Ni$(3d)$ character. 
The charge-transfer energy is computed as $\Delta_{\small\mbox{La$_2$NiO$_4$}}=3.9\,$eV, 
significantly larger than for LaNiO$_3$ and also in good accordance with experimental 
estimates~\cite{kui91}. With $\Delta\ll U$ and $\Delta\sim\Delta_g$ the system may be
classified as a charge-transfer insulator.

The LaNiO$_2$ compound with $n_d$ closer to $3d^9$ has apparently a more subtle electronic 
structure than given by an obvious metal or insulator. The Ni$(3d)$ states are nearly gapped, but
minor spectral weight remains at the Fermi level. The oxygen-hole content is non-negligible
but not as strong as for LaNiO$_3$; there are only about 0.5 holes per unit cell.
However note that the here provided fillings derived from the {\sl projected} spectral 
function may differ somewhat from values extracted from the {\sl local} spectral function.
The values $n_d^{\rm loc}$ associated with the latter function read
$\{7.95,8.15,8.84\}$ for LaNiO$_3$, La$_2$NiO$_4$ and LaNiO$_2$. The infinite-layer
compound shows the largest charge-transfer energy 
$\Delta_{\small\mbox{LaNiO$_2$}}=4.8\,$eV among the here studied nickelates.
In line with the increase of $\Delta$ with a shrinking formal oxidation state,
the O$(2p)$ levels shift to deeper energies with rising Ni$(3d)$ filling.

In the following three subsections, we will discuss the specific features of the
individual compounds in more detail.

\section{LaNiO$_3$\label{113}}
Down to low temperature, the LaNiO$_3$ compound remains in a strongly 
correlated metallic state~\cite{hor07}, with an experimental mass renormalization 
$m^*/m$ of about $3-4$~\cite{oue10,now15}. There have already been various DFT+DMFT 
accounts for this nickelate, however most studies focus on a two-orbital Ni-$e_g$
correlated subspace, and only few treat the full five-orbital Ni$(3d)$ shell as 
correlated within DMFT~\cite{den12,lia21}.

\begin{figure}[t]
\begin{center}
\includegraphics*[width=12.75cm]{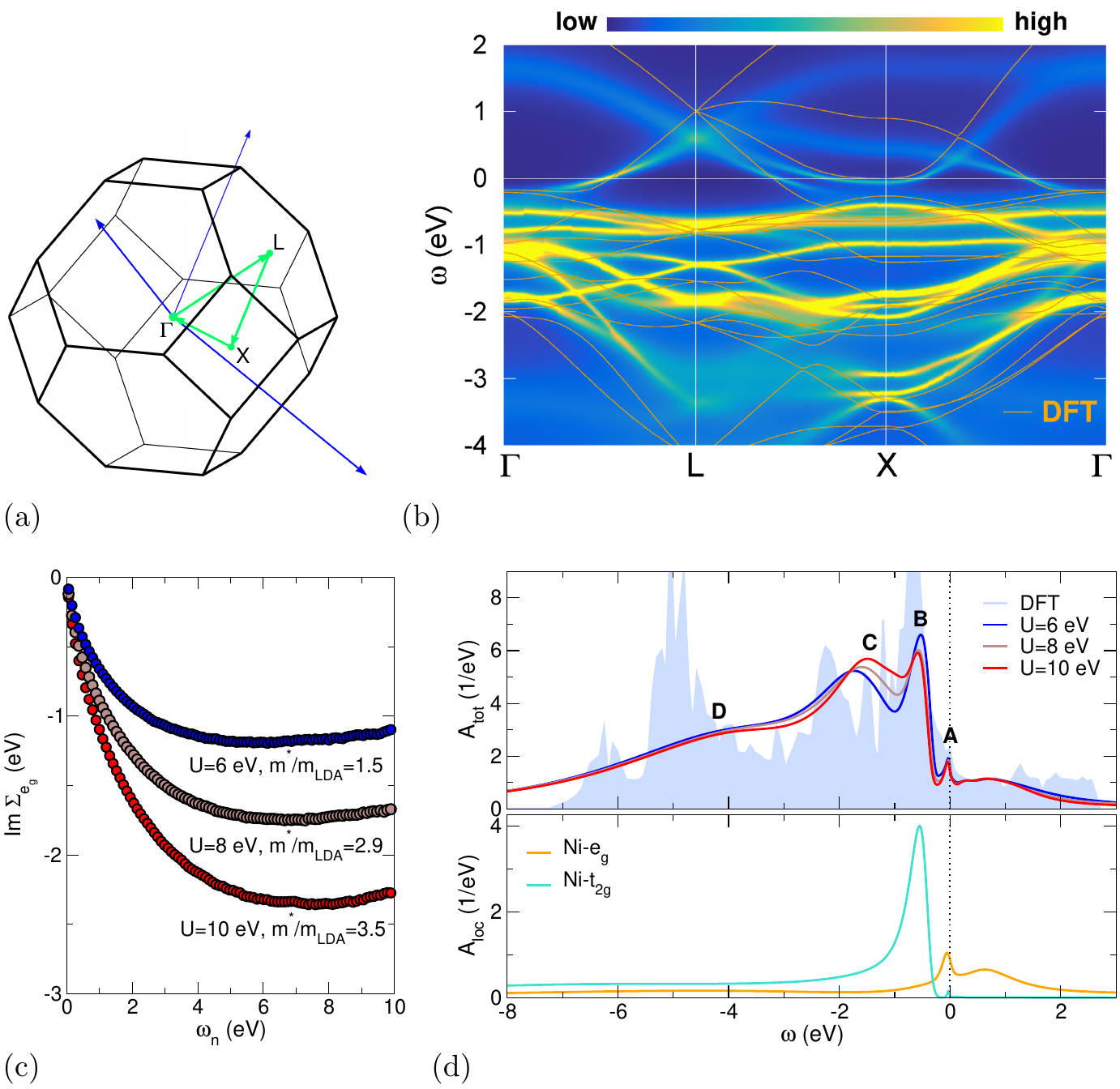}
\end{center}
\caption{Data for LaNiO$_3$.
(a) Brillouin zone with high-symmetry lines for the $R\bar{3}c$ structure.
(b-d) DFT+sicDMFT results for $T=193\,$K: (b) ${\bf k}$-resolved spectral function with
comparison to DFT dispersion; (c) imaginary part of the Ni-$e_g$ self-energy
for different $U$ values, yielding varying effective masses; (d) ${\bf k}$-integrated
total spectrum (top) and local Ni$(3d)$ spectrum (bottom).\label{fig2}}
\end{figure}

Based on our calculations utilizing the five-orbital correlated subspace, 
Fig.~\ref{fig2}b displays the {\bf k}-resolved spectral function along high-symmetry
lines in the Brillouin zone of the $R\bar{3}c$ structure (see Fig.~\ref{fig2}a). 
The Ni-$e_g$ orbitals are degenerate and two corresponding bands cross the Fermi level,
showing a DFT bandwidth of $\sim 2.5$\,eV. 
The Ni-$t_{2g}$ orbitals are mostly filled and the associated bands remain below 
$\varepsilon_{\rm F}^{\hfill}$~\cite{den12}. Compared to the DFT dispersion, there is 
a significant band renormalization for the low-energy Ni-$e_g$ dominated bands.
Figure~\ref{fig2}c documents the $U$ dependence of the local Ni-$e_g$ self-energy part 
${\rm Im}\,\Sigma(i\omega_n)$ for small Matsubara frequencies $\omega_n$, with the
extracted 
$m^*/m_{\rm LDA}=(1-\left.\partial\,\rm Im\,\Sigma/\partial\omega_n\right|_{\omega_n\rightarrow 0})^{-1}$ in this orbital basis. It is seen that the effective mass is indeed closest 
to the experimental regime for our general choice $U=10$\,eV, while it remains too small 
for a lower on-site Coulomb interaction. Note that the acutal effective mass in the band
basis, which strictly speaking should be the one for comparison with experiment, is
lower and amounts to $\left.m^*/m_{\rm LDA}\right._{\rm band}\sim 2.5$. Thus,
even a rather large $U$ value does not quite lead to a perfect-matching magnitude for the 
experimental Ni-$e_g$ band renormalization. For instance, non-local spin fluctuations
could be a further source for additional renormalization effects.

The explicit local Ni-$e_g$ occupation amounts exactly to two electrons, i.e. both the Ni-$d_{z^2}$ 
and Ni-$d_{x^2-y^2}$ orbital each host one electron. This becomes also clear from the
local spectral function shown in the bottom panel of Fig.~\ref{fig2}, where the 
half-filled $e_g$ states give rise to a renormalized quasiparticle peak (A) at the Fermi
level. The total $A(\omega)$ exhibits further peaks at $-0.6$ eV (B), $-1.5$ eV (C) and
$-4.2$ eV (D) in the occupied part of the spectrum. Peak B is associated with Ni-$t_{2g}$,
C with with O$(2p)$ and D with O$(2p)$/Ni$(3d)$-Hubbard-band. The prominent two-peak
O$(2p)$ structure at higher energies below the dominant Ni$(3d)$ part, supposedly
due to the intricate $3d^8\underline{L}$ ground state, can already be drawn from 
Fig.~\ref{fig1}. Though there is overall agreement with photoemission data~\cite{hor07}, 
there are slight differences for the peak positions; in experiment, especially peak C 
is located at deeper $-2.4$ eV. Lowering the $U$ value shifts C to somewhat to 
higher negative energies, but still not reaching the experimental location. But as
discussed before, the quasiparticle $e_g$ renormalization becomes too small for smaller
$U$. Notably, the plain DFT spectrum places the main part of peak C in better accordance
with experiment. In summary, DFT+sicDMFT provides a good description of LaNiO$_3$ for
the chosen set of Coulomb parameters, however, some questions remain concerning 
correlation strength and the general electronic spectrum. Albeit originally-thought a 
rather unspectacular compound, a very good description/understanding of the LaNiO$_3$ 
electronic structure turns out still challenging~\cite{kar19}. 

\section{La$_2$NiO$_4$\label{214}}
As the natural nickelate analog to the canonical high-$T_{\rm c}$ cuprate La$_2$CuO$_4$,
experimentally, the La$_2$NiO$_4$ compound has been studied extensively in the past (see 
e.g.~\cite{ima98} and references therein for the phenomenology with hole doping). 
However to our surprise, while there are standard-DFT~\cite{guo88,zho09}, 
DFT+U~\cite{pard12} and meta generalized-gradient approximation~\cite{lane20} studies,
we could not find an in-depth realistic-DMFT investigation for this material in the literature. 
This section is hence also devoted to fill this void.

From an experimental point of view, ensuring the exact stoichiometry for the undoped
compound, especially in view of the oxygen content~\cite{jor89,bas94}, appears the 
most critical issue in defining the correct electronic phases. Here, we do not delve into
the intriguing question of defect behavior at nominal stoichiometry, but discuss
the perfectly-ordered system.
\begin{figure}[t]
\begin{center}
\includegraphics*[width=12.75cm]{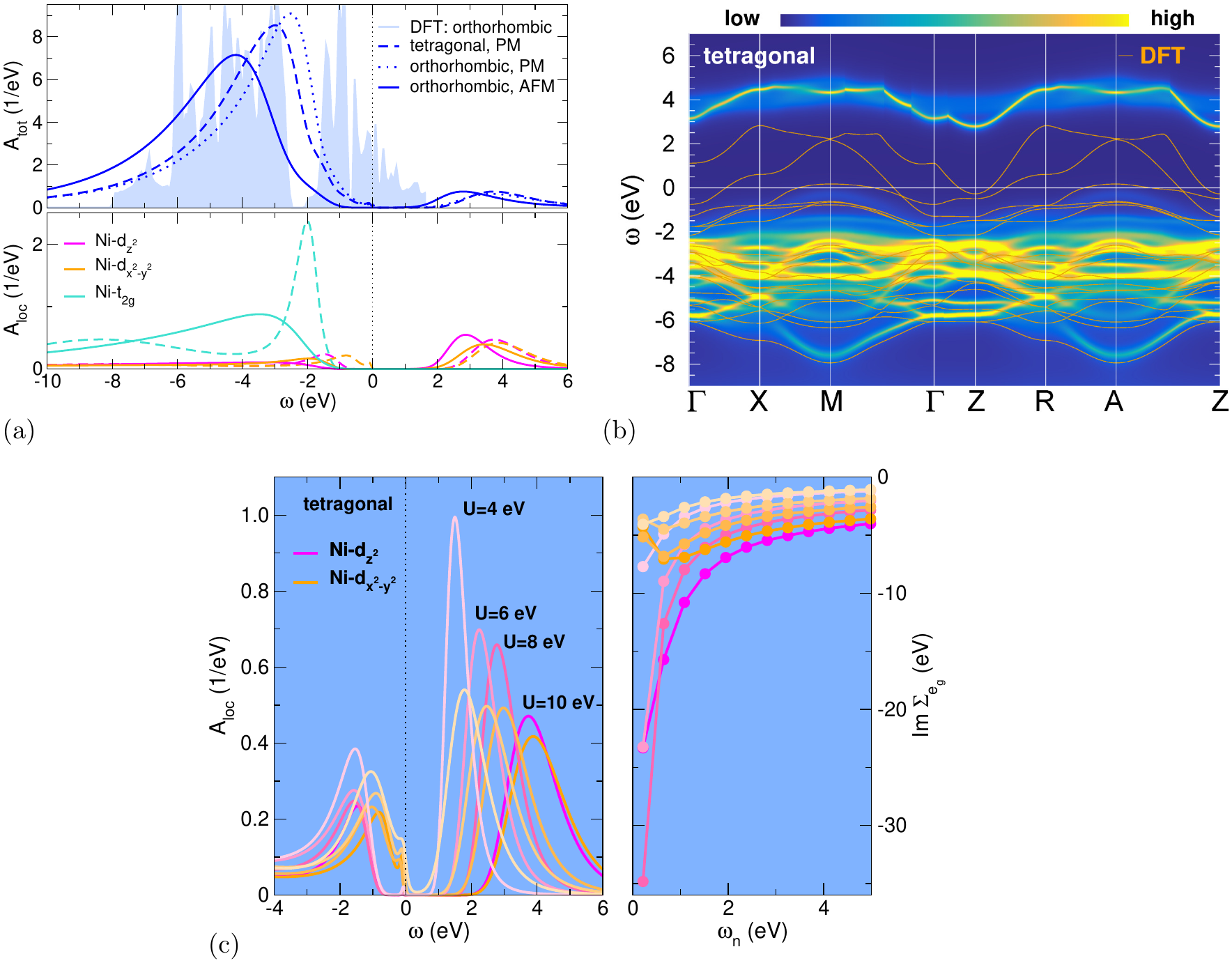}
\end{center}
\caption{DFT+sicDMFT data for La$_2$NiO$_4$.
(a) ${\bf k}$-integrated total spectrum (top) and local Ni$(3d)$ spectrum (bottom)
for different structural and magnetic cases at $T=193\,$K.
(b) ${\bf k}$-resolved spectral function for the tetragonal phase at $T=800$\,K 
with comparison to DFT dispersion.
(c) Local Ni-$e_g$ spectral function (left) and imaginary part of the the Ni-$e_g$
self-energies (right) for different $U$ values in the tetragonal phase at 
$T=800$\,K.\label{fig3}}
\end{figure}

The orthorhombic compound at ambient and lower temperature is described metallic in
nonmagnetic DFT (see Fig.~\ref{fig3}a). By imposing strong electronic correlations and
additional antiferromagnetic order, La$_2$NiO$_4$ is transfered into a Mott-insulating
state with a sizable charge gap of $\sim 4$\,eV. That latter value and the associated
electronic spectrum (full line in top panel of Fig.~\ref{fig3}a) are in very good 
agreement with data from photoelectron spectroscopy~\cite{eis92}. There is a prominent 
and broad peak centered at $-4$\,eV, mainly constituted by O$(2p)$ (cf. mid panel of
Fig.~\ref{fig1}b). The upper Hubbard-band edges of Ni-$e_g$ character are responsible
for a shoulder at $-2$\,eV. While the Ni-$t_{2g}$ orbitals are once more completely filled,
Ni-$d_{x^2-y^2}$ and Ni-$d_{z^2}$ both host one electron, i.e. 
$\{n_{z^2},n_{x^2-y^2}\}=\{1.07,1.09\}$, respectively. The ordered
Ni moment in the standard G-type AFM phase turns out $m_{\rm Ni}=1.65$\,$\mu_{\rm B}$
at $T=193$\,K from our IC-scheme calculations, with an equal share of spin polarization 
between both Ni-$e_g$ orbitals. Thus the Ni moment deviates significantly from the value
associated with a saturated $S=1$ spin, but is again in very good agreement with 
experimental values~\cite{lan89,rod91}. Note that from neutron diffraction the N{\'e}el 
temperature is located at $T_{\rm N}=330$\,K~\cite{rod91}. It is however important to 
realize that the magnetic order is not decisive for establishing the insulating state.

The high-temperature phase of La$_2$NiO$_4$ has been a matter of debate in the past.
Transport studies~\cite{gan73} motivated a (bad-)metallic scenario for the 
tetragonal phase of La$_2$NiO$_4$, i.e. some kind of metal-to-insulator transition with 
rising temperature. Goodenough and coworkers~\cite{goo82,fon85} suggested an 
orbital-selective Mott transition upon heating, whereby the Ni-$d_{z^2}$ states
remain insulating but the Ni-$d_{x^2-y^2}$ states become metallic. From a structural
point of view, the absence of the further extension and rotation of the NiO$_6$ octahedron 
known within the $Bmab$ phase, leads to a weaker splitting between both 
$e_g$ states~\cite{zho09}. The DFT crystal-field splitting $\Delta_{\rm cf}=
\varepsilon_{z^2}-\varepsilon_{x^2-y^2}$ amounts to 66\,meV in the orthorhombic phase
and to 38\,meV in the tetragonal phase.
Additionally, while the total Ni-$e_g$ bandwidth in the tetragonal phase amounts to $\sim 2.8$\,eV,
the individual orbital contributions differ (cf. Fig.~\ref{fig3}b). 
The $d_{x^2-y^2}$ dominated DFT band crossing the Fermi level along $\Gamma$-X is 
indeed significantly wider than the $d_{z^2}$ dominated band crossing along X-M. However,
the orbital $e_g$ mixing within these bands is still about 2:1 (and vice versa, 
respectively), due to the reduced energy splitting. As a further interesting aspect, 
the DFT band structure of tetragonal La$_2$NiO$_4$ shows self-doping behavior through a 
largely La$(5d)$ dominated band creating a small electron pocket around the Z point. This 
highlights the stronger three-dimensional character compared to e.g. La$_2$CuO$_4$.

The plotted spectral parts and self-energies in Fig.~\ref{fig3} render obvious that within 
DFT+sicDMFT there is no orbital-selective physics in the tetragonal phase at higher temperature. 
While indeed the Ni-$d_{x^2-y^2}$ orbital turns out somewhat weaker correlated than the 
Ni-$d_{z^2}$ one, the gap opening takes place in both orbital sectors also for $U<10$\,eV.
Still, the $U=10$\,eV regime appears again best fitting to La$_2$NiO$_4$, as the charge gap results
too small for reduced $U$ values. Thus the possible (bad-)metallic behavior at higher temperatures
in this compound should be associated with the intriguing defect characteristics at
stoichiometry~\cite{jor89,bas94}.

\section{LaNiO$_2$}
The infinite-layer LaNiO$_2$ belongs to the growing family of superconducting
nickelates upon thin-film growth and Sr doping~\cite{li19,osa21,zen21}. This finding
has stimulated a vast theory interest in these type of compounds 
(see e.g.~\cite{wu19,nom19,hu19,jia19,jiang19,wer20,li20,guzhu20,leo20,wankang20,hep20,zhav20,bot20,si20,ole20,cho20-2,kar20-1,zhalan21,kat20,bee21,pli21,cho21,kar21,lec20-1,lec20-2,lec21}
among others), with notable earlier work~\cite{ani99,lee04}.
For a detailed discussion within the DFT+sicDMFT approach for the case of NdNiO$_2$, 
we refer to our previous works~\cite{lec20-1,lec20-2,lec21}.

\begin{figure}[t]
\begin{center}
\includegraphics*[width=12.75cm]{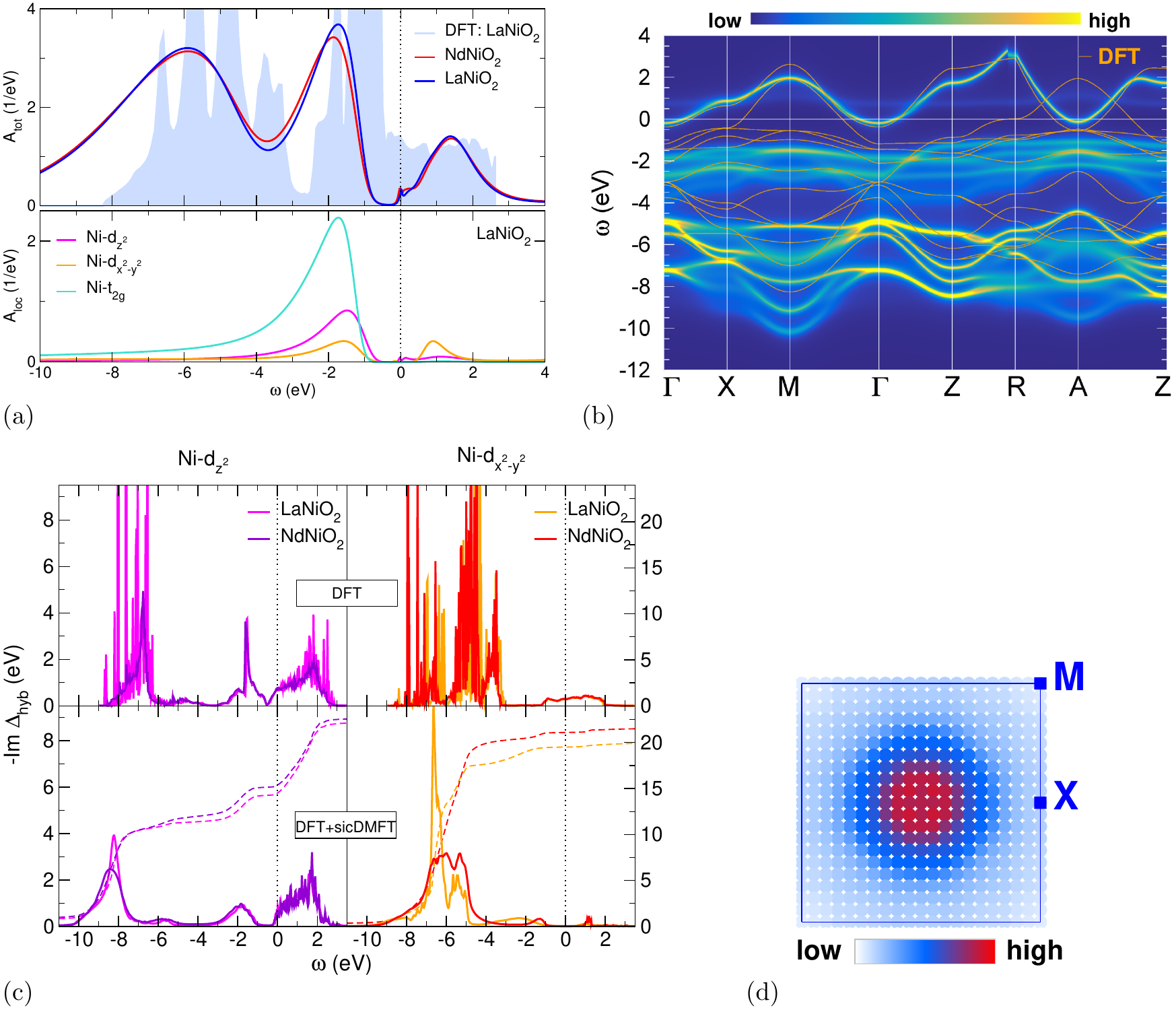}
\end{center}
\caption{DFT+sicDMFT data for LaNiO$_2$ ($T=193\,$K).
(a) ${\bf k}$-integrated total spectrum (top), comparing with the NdNiO$_2$ total 
spectrum, and local Ni$(3d)$ spectrum (bottom).
(b) ${\bf k}$-resolved spectral function with comparison to DFT dispersion.
(c) Imaginary part of the Ni-$e_g$ hybridization function displayed as 
$-{\rm Im}\,\Delta_{\rm hyb}$ in the DFT limit (top) and in the interacting regime
(bottom). (d) Static weak-couling spin susceptibility $\chi_s({\bf q},0)$ for 
${\bf q}=(q_x,q_y,0)$.\label{fig4}}
\end{figure}

Non-surprisingly, the essential physics of LaNiO$_2$ far away from the low-temperature
regime turns out very similar to NdNiO$_2$ on the present level of theory. 
Figure~\ref{fig4}a-c display the key features of the
paramagnetic electronic spectrum. The DFT density of states describes a good metallic
system, yet strong electronic correlations transfer most of the low-energy spectral
weight to higher energies. Note that the DFT+sicDMFT predicted spectrum agrees well
with data from recent photoemission spectroscopy performed on thin films of 
PrNiO$_2$~\cite{chen21}, where the two main peaks at $-2$\,eV and $-5.5$\,eV are well
confirmed by experiment.
The Ni$^+(3d^9)$ configuration leads to an intriguing situation 
where the Ni-$d_{x^2-y^2}$ orbital is close to half filling and residual
hole character in the Ni-$d_{z^2}$ orbital remains (cf. Fig.~\ref{fig4}a, lower panel). 
That latter character is achieved via a self-doping band of dominant La$(5d)$ nature, 
causing electron pockets around the $\Gamma$ and A point in reciprocal space, 
which still carry minor Ni-$d_{z^2}$ hybridization weight. Contrary to the self-doping band in 
tetragonal La$_2$NiO$_4$, the electron pockets at the Fermi level survive the impact of strong 
electronic correlations. Reason is that those pockets do not carry Ni-$d_{x^2-y^2}$ weight and
are therefore not shifted to higher energies. Thus the effective Ni-$d_{x^2-y^2}$
band is most-proximate to a (orbital-selective) Mott transition and the largely-filled 
and therefore weakly-to-moderately correlated Ni-$d_{z^2}$ orbital remains in a joint 
metallic state with the La$(5d)$ orbitals (see Fig.~\ref{fig4}b). 
For illustration, we also show in Fig.~\ref{fig4}c the imaginary diagonal part of the Ni-$e_g$
hybridization function $\Delta_{\rm hyb}$ in the DFT limit~\cite{kar21} as well as in
the interacting regime with 
$\Delta_{\rm hyb}(\omega)=\omega+\mu-G_{\rm loc}^{-1}(\omega)-\Sigma_{\rm imp}(\omega)$.
In the DFT limit, both Ni-$e_g$ have sizable weight at low-energy, whereas the hybridization
function for the Ni-$d_{x^2-y^2}$ vanishes in DFT+sicDMFT close to the Fermi level because
of its Mott-insulating nature. The hybdrization structures with O$(2p)$ are dominant at
higher energy around $-6$\,eV especially for Ni-$d_{x^2-y^2}$. Note the energy shift towards
deeper energies due to the SIC inclusion compared to standard DFT. Comparing LaNiO$_2$ and
NdNiO$_2$ does not reveal striking differences in $\Delta_{\rm hyb}$. But it is revealed
that the frequency-integrated hybridization (dashed lines in lower panel of Fig.~\ref{fig4}c), 
is somewhat stronger for the Nd compound. This could affect the low-temperature physics in 
some regard, and different features seem indeed revealed experimentally~\cite{osa21}
(see also text below).

Loosely speaking, infinite-layer nickelate with the missing apical oxygen can be viewed 
as a coupling of two dichotomic subsystems: strongly hybridized/correlated 
Ni-$d_{x^2-y^2}$$-$O$(2p)$ and weakly-correlated Ni-$d_{z^2}-$RE$(5d)$.
Hole doping leads to a further rise of complexity in the entanglement of these two 
subsystems, namely a change from (orbital-selective) Mottness to Hundness via the
enclosed superconducting region~\cite{lec20-1,lec20-2,lec21}. But we will not pursue 
this physics in the present context.

Instead, we here want to briefly comment on the issue of ordered magnetism, since the spin physics 
of RENiO$_2$ compounds is highly puzzling in experiment. Absence of long-range 
order~\cite{hay99,wang20}, quasi-static ordering~\cite{cui20}, spin-glass physics~\cite{lin21}, 
pseudogap vs. Curie-Weiss~\cite{zhao21} and (para)magnon dispersions~\cite{lu20} are discussed.
Various theoretical accounts~\cite{leo20-2,zhalan21,kat20} find a nearest-neighbor superexchange 
coupling $J\sim 60-70$\,meV, i.e. about half the size of the value in high-$T_{\rm c}$ cuprates, and 
antiferromagnetic order should thus be favorable. However one must also remember that all available 
calculations find a metallic state for RENiO$_2$ at stoichiometry, either due to a robust Ni-$d_{x^2-y^2}$
quasiparticle weight and/or, as here, because of the existent self-doping band. This is in
stark contrast to layered cuprates. But metallicity,
strong electronic correlations and AFM order do not go along very well in nature, reason why
there are only few reported materials cases (see e.g. discussion in~\cite{kom11}). 
Furthermore, the stronger three-dimensional character of RENiO$_2$ compared to layered cuprates 
does not help either to stabilize AFM order under these circumstances. Therefore given the 
electronic structure of the infinite-layer nickelates, the absence of robust AFM order in 
experiment is in fact not totally surprising. To underline this viewpoint, we show in 
Fig.~\ref{fig4}d the diagonal part of the LaNiO$_2$ static weak-coupling spin susceptibility 
$\chi_s({\bf q},0)=-\frac{1}{\beta}\sum_{{\bf k}n,\nu\nu'}G_{\nu}({\bf k},i\omega_n)G_{\nu'}({\bf k}+{\bf q},i\omega_n)$, where $\omega_n$ are fermionic Matsubara frequencies and $G_\nu$ denotes the converged
DFT+sicDMFT Green's function in Bloch space for band indeces $\nu\nu'$. 
This simplified Lindhard-like susceptibility neglects 
vertex contributions and makes only sense in a system with existing low-energy spectral weight. 
Because of the self-doping electron pockets around $\Gamma$ and A (plus some very weak leftovers from
the effectively Mott-insulating Ni-$d_{x^2-y^2}$ band), $\chi_s$ is nonzero and obviously has
highest intensity around $\Gamma$. Note that the absolute values are still low and far off any
instability regime. Still, the maximum close to the zone centre puts the system in favor of 
ferromagnetic ordering tendencies from this perspective. These tendencies oppose the 
strong-coupling AFM tendencies from the superexchange perspective. Again to be clear, we do not 
at all claim that the given $\chi_s$ describes the full range of spin fluctuations in LaNiO$_2$ 
adequately, it only shines a light on the implications coming from the metallic side of the problem.
Let us also mention that additionally, a unique Kondo scenario involving 
the Ni-$d_{x^2-y^2}$ spin and the self-doping band has been detected below 
$T\sim 60$\,K for the case of NdNiO$_2$~\cite{lec20-2}. 
Note that again the Ni-$d_{z^2}$ orbital, hybridized onto the 
self-doping band, plays a decisive role in arranging for link between Ni-$d_{x^2-y^2}$ spin and 
itinerant electrons. It was shown~\cite{lec21}, that this Kondo coupling may interfere/coexist 
with AFM ordering tendencies at low temperatures.

Equipped with our IC scheme to handle finite spin polarization in the interacting regime, the
DFT+sicDMFT calculations for most promising C-type AFM order~\cite{cho20-2,zhalan21,lec20-2}, 
i.e. AFM in-plane and FM out-of-plane, results in ordered Ni moments of size 
$m_{\rm Ni}\leq 0.05$\,$\mu_{\rm B}$. This result holds for $T=190$\,K as well as for
the lower-temperature regime of $T=30$\,K. Thus in qualitative agreement with our previous
work on NdNiO$_2$, the tendency to establish sizable-moment antiferromagnetism in RENiO$_2$
is weak, in agreement with experimental findings. 
Note that we employed identical initialization protocols with starting moments to
eventually stabilize magnetic order, and while the $d^8$ compound La$_2$NiO$_4$ develops
substantial Ni moments (see previous section), the $d^9$ compound LaNiO$_2$ does not.
However, the residual magnetic order apparently helps to establish a stable electronic
LaNiO$_2$ phase at lower temperature $T=30$\,K. In contrast to NdNiO$_2$~\cite{lec20-2,lec21}, 
the Ni-$d_{z^2}$ orbital in the PM phase seemingly becomes in some way critical at lower $T$
as the associated self-energy develops unphysical features at small Matsubara frequencies.
This might be a numerical artifact, but we still want to report it here, since several
careful attempts to cure these features were unsucessful. One then surely enters highly
speculative territory, but the basic message could be that the Kondo-decorated PM phase
(and its underlying unconventional Ni-$e_g-$RE$(5d)$ coupling structure) detected for
NdNiO$_2$, is in some way 'disturbed' in LaNiO$_2$ at comparable temperatures.

\section{Conclusions}
In this work we presented a comparison of the correlated electronic structure of three
different $3d^n$ nickelates, i.e., formal-$d^7$ LaNiO$_3$, $d^8$ La$_2$NiO$_4$ and
$d^9$ LaNiO$_2$, based on calculations within the DFT+sicDMFT scheme. One goal was to
evaluate the performance of the latter scheme for the wider class of nickel-oxide compounds.
Moreover, we chose to fix the local Coulomb
parameters $U$ and $J_{\rm H}$ to examine if key features of the different compounds
can still be well described without fine tuning (or direct calculation) of the interaction 
integrals. Concerning both aspects, positive results are reported here. The DFT+sicDMFT approach 
is very capable of describing the correlated-metal nature of LaNiO$_3$ with its hole on 
oxygen as well as the Mott-insulating characteristics of La$_2$NiO$_4$. And importantly,
the constant-interaction strength approach shows that while a detailed knowledge/calculation
of the Hubbard parameters is surely a relevant goal, main electronic-structure features
are well describable without strong changes of the interaction strength from one compound
to another. For the case of LaNiO$_2$, comparison to experiment remains still difficult due
to the scarce availability of measured data. But the present results for the other two 
compounds and recent photoemission data~\cite{chen21} provide some confidence that the 
here (and elsewhere~\cite{lec20-1,lec20-2,lec21}) established picture for infinite-layer 
nickelates will also prove reliable.

In the course of this materials study, a refined scheme to perform spin-polarized DFT+(sic)DMFT
studies was introduced. This intermediate-coupling scheme spin averages the DFT charge
density terms that connect to the local part of the problem, but still allows for finite spin
polarization in the plane-wave-derived charge density within DFT. Hence most of the 
spin polarization should be generated in the DMFT part, yet a minor feedback into the DFT part
is restored. This new scheme provides ordered Ni moments for La$_2$NiO$_4$ in excellent agreement
with experiment. It furthermore confirms our recent results~\cite{lec21} of a (nearly) 
vanishing AFM-ordered Ni moment in infinite-layer nickelates, also in line with experiment. 
The intriguing metallic(-like) nature at stoichiometry may be blamed for prohibiting a strongly 
correlated AFM state.

Finally, let us reiterate on the fact that nickelates, in general, are manifest Ni-$e_g$ systems.
While for LaNiO$_3$, the $d_{z^2}$ and $d_{x^2-y^2}$ orbital act quite coherently, some dichotomy
sets in for La$_2$NiO$_4$. Crystal-field splitting and different bandwidths cause orbital
differentiation for this layered $d^8$ compound, but in the end both orbital sectors give rise
to similar qualitative behavior. Then for LaNiO$_2$ this dichotomy is forced to an utmost limit,
where half-filled $d_{x^2-y^2}$ is (nearly) Mott-insulating and largely-filled $d_{z^2}$ remains 
metallic. It is interesting to note that, as shown, orbital selectivity and self-doping character 
are already a matter of debate for La$_2$NiO$_4$, namely in its high-temperature tetragonal
phase. However, these issues apparently only become severe as a 'game changer' for the $d^9$
LaNiO$_2$. In this context it may also be worth to state that the mechanism of getting rid of
the DFT self-doping band via rotation/distortion in the orthorhombic phase of La$_2$NiO$_4$,
has just been shown to may also be a proper mechanism for later RENiO$_2$ 
systems~\cite{xia21,ber21}. It still has to be explored if the even more extreme dichotomic limit
of being able to abandon one of the Ni-$e_g$ orbitals completely from a low-energy discussion
can eventually be realized. In any case, the plethora of puzzling and demanding physics that emerges 
from the Ni-$e_g$ manifold in nickel oxides will remain an exciting research area in condensed 
matter physics.

\ack
The author is grateful to A. J. Millis for helpful discussions. 
Calculations were performed at the Juwels Cluster of the J\"ulich Supercomputing Centre 
(JSC) under the hhh08 project.

\section*{References}
\bibliographystyle{iopart-num}
\bibliography{bibscan}

\end{document}